\documentclass[3p,times,procedia]{elsarticle}
\usepackage{nupha_ecrc}

\volume{00}

\firstpage{1}

\journalname{Nuclear Physics A}

\runauth{D. Bollweg et al.}

\jid{nupha}

\jnltitlelogo{Nuclear Physics A}

\usepackage{amssymb}
\usepackage{amsthm}
\usepackage{amsmath}
\usepackage{lineno}

\usepackage[figuresright]{rotating}

\begin{document}

\begin{frontmatter}



%% Instructions from Editor: Please use the following \dochead only in the preprint version (e-print arXiv etc.); 
%% use empty \dochead{} when submitting to Nuclear Physics A!
\dochead{XXVIIIth International Conference on Ultrarelativistic Nucleus-Nucleus Collisions\\ (Quark Matter 2019)}
%% e.g. \dochead{17th International Conference on Dynamical Processes in Excited States of Solids}

\title{Higher order cumulants of net baryon-number distributions at non-zero $\mu_B$}

\author[Bielefeld]{D. Bollweg}
\author[Bielefeld]{F. Karsch}
\author[BNL]{S. Mukherjee}
\author[Bielefeld]{C. Schmidt}
\address[Bielefeld]{Fakult\"at f\"ur Physik, Universit\"at Bielefeld, D-33615 Bielefeld, Germany}
\address[BNL]{Physics Department, Brookhaven National Laboratory, Upton, New York 11973, USA}

\begin{abstract}
Using recent results on higher order cumulants of conserved charge fluctuations from lattice QCD, we construct mean, variance, skewness, kurtosis, hyper-skewness and hyper-kurtosis of net-baryon number distributions for small baryon chemical potentials $\mu_B$. For the strangeness neutral case (\(\mu_S=0\)) at fixed ratio of electric charge to baryon number density (\(\frac{n_Q}{n_B}=0.4\)), which is appropriate for a comparison with heavy ion collisions, we present results for \(\kappa_B \sigma_B^2\), \(S_B \sigma_B^3/M_B\),  \(\kappa^{H}_{B}\sigma_{B}^4\) and \(S^{H}_{B}\sigma^5_{B}/M_{B}\) on the crossover line for the chiral transition, \(T_{pc}(\mu_B)\). Continuum extrapolations for this pseudo-critical transition line have recently been reported by HotQCD up to baryon chemical potentials \(\mu_B\simeq 300\) MeV \cite{Bazavov_2019}. These cumulant ratios are of direct relevance for comparisons with corresponding ratios measured by STAR in the BES-I and II runs at beam energies \(\sqrt{s_{NN}}\ge 20\) GeV. In particular, we point out that recent high statistics results on skewness and kurtosis of net-baryon number distributions obtained by STAR at  \(\sqrt{s_{NN}} = 54.4\) GeV put strong constraints on freeze-out parameters and are consistent with predictions from thermal QCD.

\end{abstract}


\end{frontmatter}

%%\linenumbers

\section{Introduction}
\label{Introduction}
Fluctuations of conserved charges such as strangeness, baryon-number and electric charge serve as ideal probes to study the phase structure of QCD because higher order cumulants of these fluctuations diverge at a critical point if they couple to the order parameter. Via measurements of event-by-event fluctuations of associated proxy particle species, these conserved charge fluctuations are accessible to heavy ion collision experiments. Lattice QCD predictions of such fluctuations remain difficult since direct simulations at finite baryon chemical potential $\mu_B$ are hampered by the infamous sign problem. Instead, numerically expensive extrapolation methods have to be used. Here we employ Taylor-expansions around $\mu_B=0$ to obtain cumulants of net baryon-number distributions at small $\mu_B$ with the aim to provide first-principle thermal QCD baselines for comparison with recent measurements of net proton-number cumulants performed by STAR.

\section{Numerical Setup}
\label{Numerical Setup}
We perform rational hybrid Monte Carlo simulations of (2+1)-flavor highly improved staggered quarks with quark masses tuned to the physical point. Compared to previous studies by HotQCD \cite{Bazavov_2017}, the number of gauge configurations in the vicinity of $T_{pc}$ is increased by a factor of three to five for lattices with temporal extend $N_{\tau}=8$ and aspect ratio $N_{\sigma}/N_{\tau}=4$. For lattices with $N_{\tau}=12$ the number of gauge configurations is increased by a factor of six to eight. The exact numbers of configurations per temperature and temporal extent are given in Table 1 of \cite{bazavov2020skewness}. Using this high statistics data set, we are able to compute generalised susceptibilities $\chi^{BQS}_{ijk}$ up to order $i+j+k=8$:
\begin{align}
    \chi^{BQS}_{ijk}(T,\vec{\mu})=\frac{1}{VT^3}\frac{\partial^{i+j+k}\log{Z(T,\vec{\mu})}}{\partial\hat{\mu}_B^i\partial\hat{\mu}_Q^j\partial\hat{\mu}_S^k}, \;\; \hat{\mu}_X\equiv\frac{\mu_X}{T}.
\end{align}
These susceptibilities constitute Taylor expansion coefficients of higher order cumulants of conserved charge distributions.
\begin{align}
  \chi^{X}_{n}(T,\mu_B)=\sum_{k=0}^{k_{\mathrm{max}}}\tilde{\chi}_{n}^{X,k}(T)\hat{\mu}_{B}^{k}, \;\;\mathrm{with}\;\; X=B,Q,S
\end{align}
The expansion is constrained such that the isospin imbalance ($n_Q/n_B=0.4$) and strangeness neutrality ($n_S=0$) conditions, similar to those relevant in heavy ion collisions, are
fulfilled order by order. For the case of net baryon-number flucutations, the explicit form of the coefficients \(\tilde{\chi}_{n}^{B,k}\) can be found in \cite{bazavov2020skewness}. With this setup, first and second order cumulants, related to mean and variance, respectively, are computed to NNNLO, whereas third and fourth order cumulants are obtained to NNLO and fifth and sixth order cumulants to NLO in $\mu_B/T$. We use them to form cumulant ratios related to those measured in heavy ion collision experiments
\begin{align}
  \label{eq:rdef}
  R^{X}_{nm}=\frac{\chi^{X}_{n}(T,\mu_B)}{\chi^{X}_{m}(T,\mu_B)}=\frac{\sum_{k=0}^{k_{\mathrm{max}}}\tilde{\chi}_{n}^{X,k}(T)\hat{\mu}_{B}^{k}}{\sum_{l=0}^{l_{\mathrm{max}}}\tilde{\chi}_{m}^{X,l}(T)\hat{\mu}_{B}^{l}}.
\end{align}

These ratios stemming from lattices with different $N_{\tau}$ are jointly fitted using a rational polynomial ansatz that includes $1/N_{\tau}^2$ corrections to obtain continuum estimates, see \cite{bazavov2020skewness} for further details.

\section{Results}
\begin{figure}[h]
  \centering
  
  \includegraphics[width=.49\textwidth]{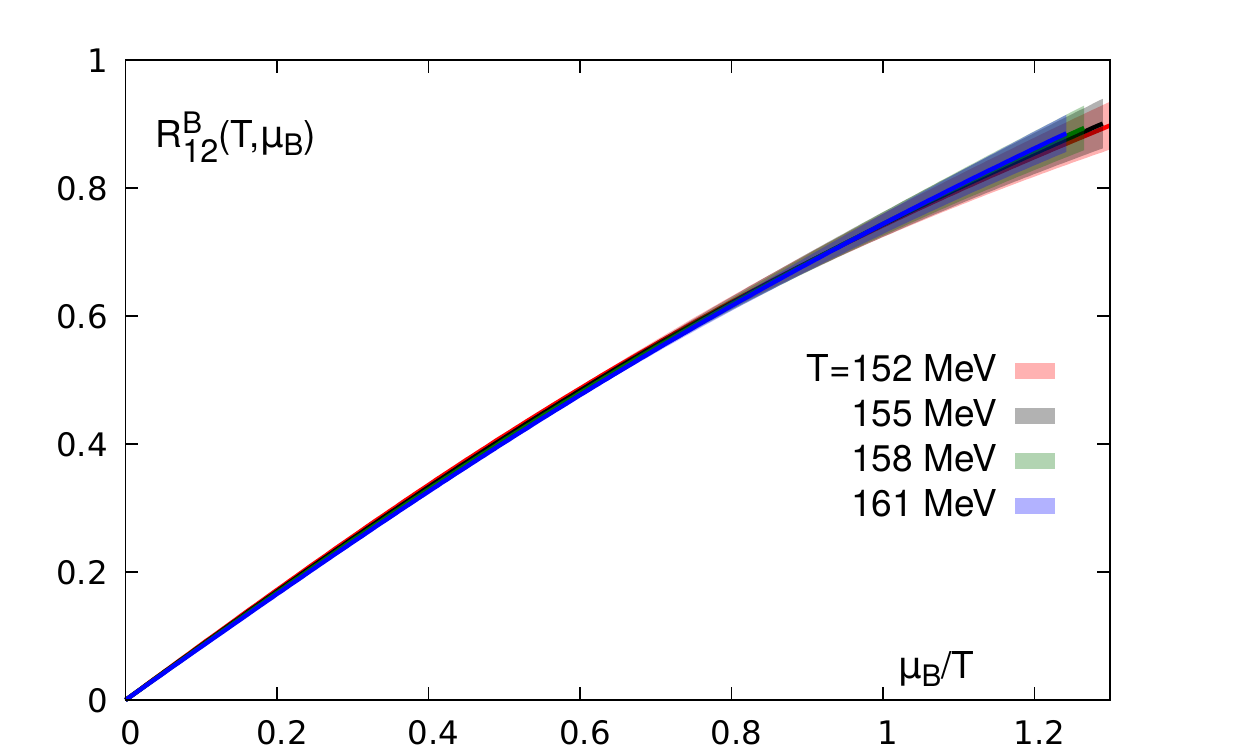}
  \includegraphics[width=.49\textwidth]{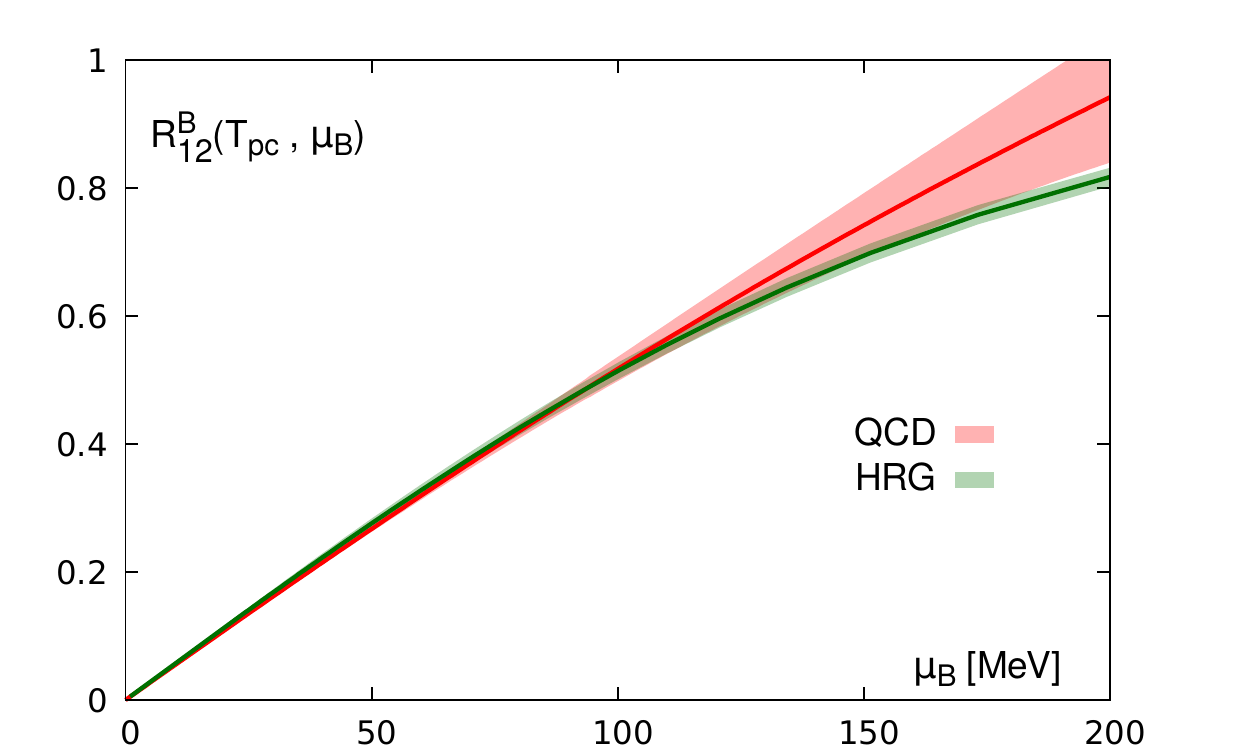}
  
  \caption{Left figure: Temperature variation of $R^{B}_{12}(T,\mu_B)$. Right figure: Comparison of $R^{B}_{12}(T_{\mathrm{pc}},\mu_B)$ from lattice QCD and HRG.} 
  \label{fig:r12b}
\end{figure}

\begin{figure}[h]
  \centering
  \includegraphics[width=.49\textwidth]{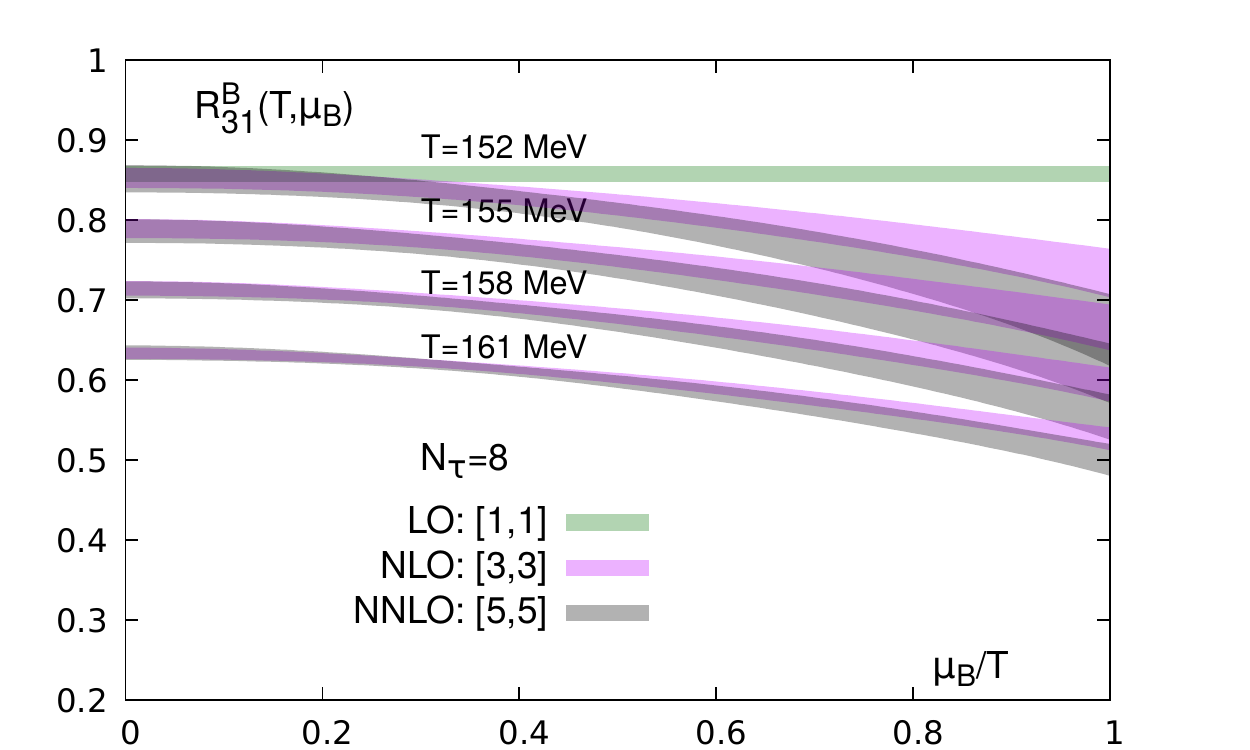}
  \includegraphics[width=.49\textwidth]{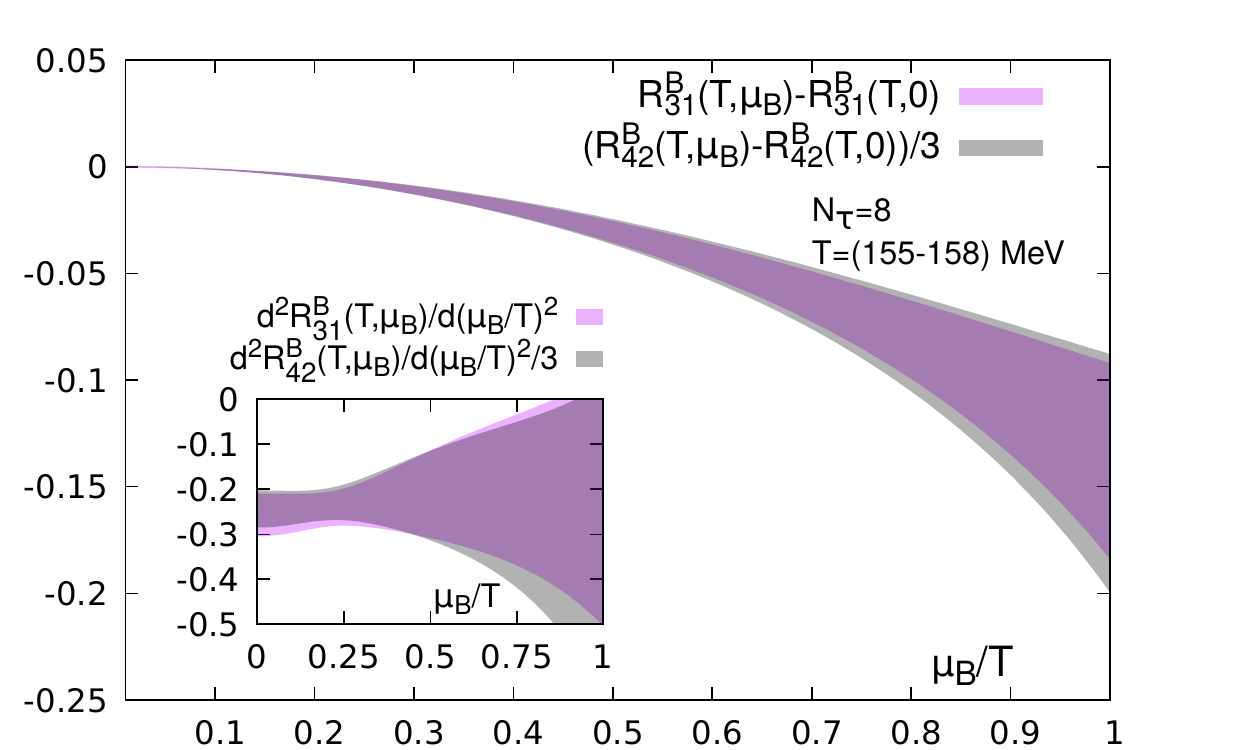}
  \caption{Left figure: Temperature variation of $R^{B}_{31}(T,\mu_B)$ for different combinations of $[k_{\mathrm{max}},l_{\mathrm{max}}]$, see eq. \eqref{eq:rdef}. Right figure: Comparison between the $\mu_B$-dependent parts of $R^{B}_{31}$ and $R^{B}_{42}/3$, with their second derivatives shown in the inset.}
  \label{fig:r3142}
\end{figure}

In Fig.\ref{fig:r12b} (left) we show the continuum extrapolated ratio of mean and variance of the net baryon-number distribution \(R^{B}_{12}(T,\mu_B)=M_{B}/\sigma^2_{B}\)
obtained from our lattice simulations for temperatures around $T_{pc}(\mu_B=0)$. We find the temperature variation to be rather weak and the $\mu_B$-dependence to be dominated by the linear, leading order contribution. Evaluating the ratio of mean and variance along the pseudo-critical transition line $T_{\mathrm{pc}}(\mu_B)=T_{\mathrm{pc}}^{0}(1+\kappa^{B}_{2}\mu_B^2)$ reported in \cite{Bazavov_2019} and comparing it to Hadron-Resonance-Gas (HRG) models, shown in Fig.\ref{fig:r12b} (right), we find the deviations from the leading order linear behaviour to be weaker than predicted by HRG models. Up to about 125 MeV, the HRG models agree with our lattice calculations but underestimate them when approaching larger values of $\mu_B$.\\
Furthermore, we calculate the skewness and kurtosis ratios of the net baryon-number distribution,
\begin{align}
  R^{B}_{31}(T,\mu_B)&=\frac{S_{\!\!B}\sigma^3_{B}}{M_B}=\frac{\chi^{B}_{3}(T,\mu_B)}{\chi^{B}_{1}(T,\mu_B)}\;\; \mathrm{and}\;\;  R^{B}_{42}(T,\mu_B)=\kappa_{B}\sigma^2_{B}=\frac{\chi^{B}_{4}(T,\mu_B)}{\chi^{B}_{2}(T,\mu_B)}.
\end{align}
The leading order contributions to these ratios are independent of $\mu_B$, while NLO contributions are slightly smaller than zero, giving rise to a weaker $\mu_B$-dependence for these ratios than $R^{B}_{12}$. Instead, they show a strong variation with temperature as exemplified by $R^{B}_{31}$ in Fig.\ref{fig:r3142} (left). In \cite{Bazavov_2017} it was found that the curvature of the kurtosis ratio was about a factor three larger than that of the skewness ratio. As shown in Fig.\ref{fig:r3142} (right) this remains true even with the inclusion of NNLO contributions.

\begin{figure}[h]
  \centering
  \includegraphics[width=.49\textwidth]{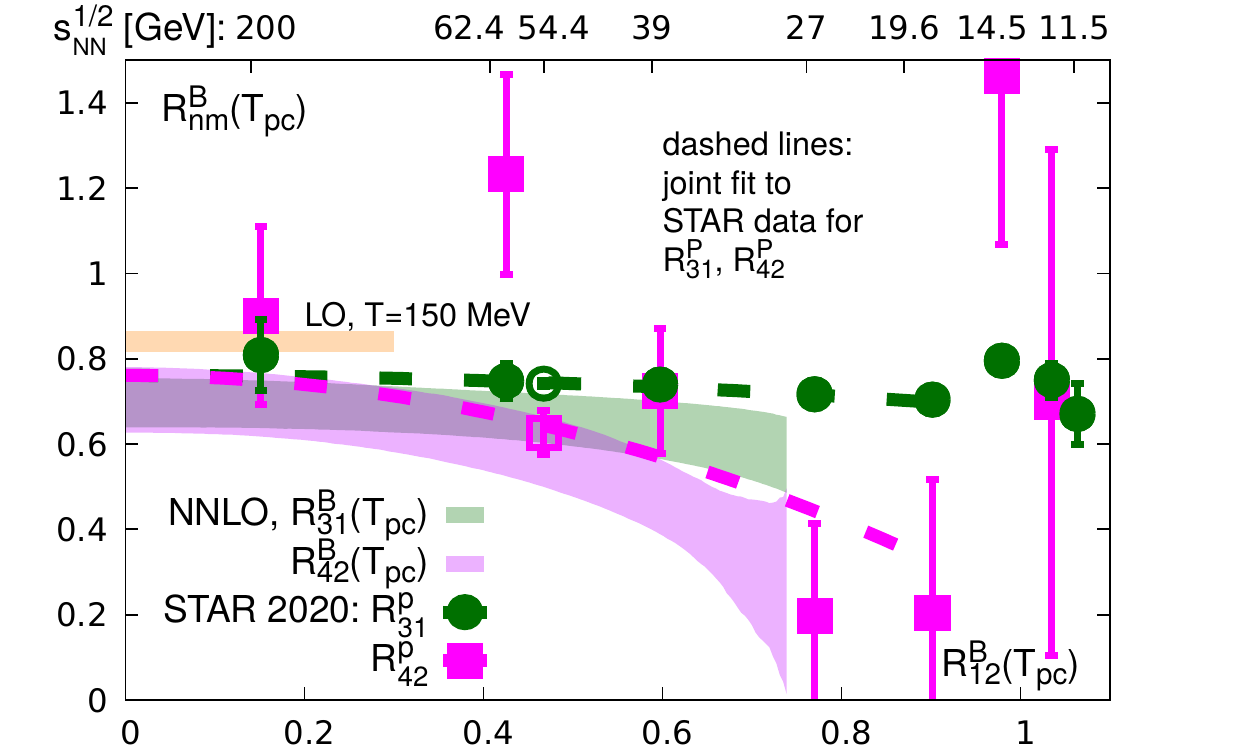}
  \includegraphics[width=.49\textwidth]{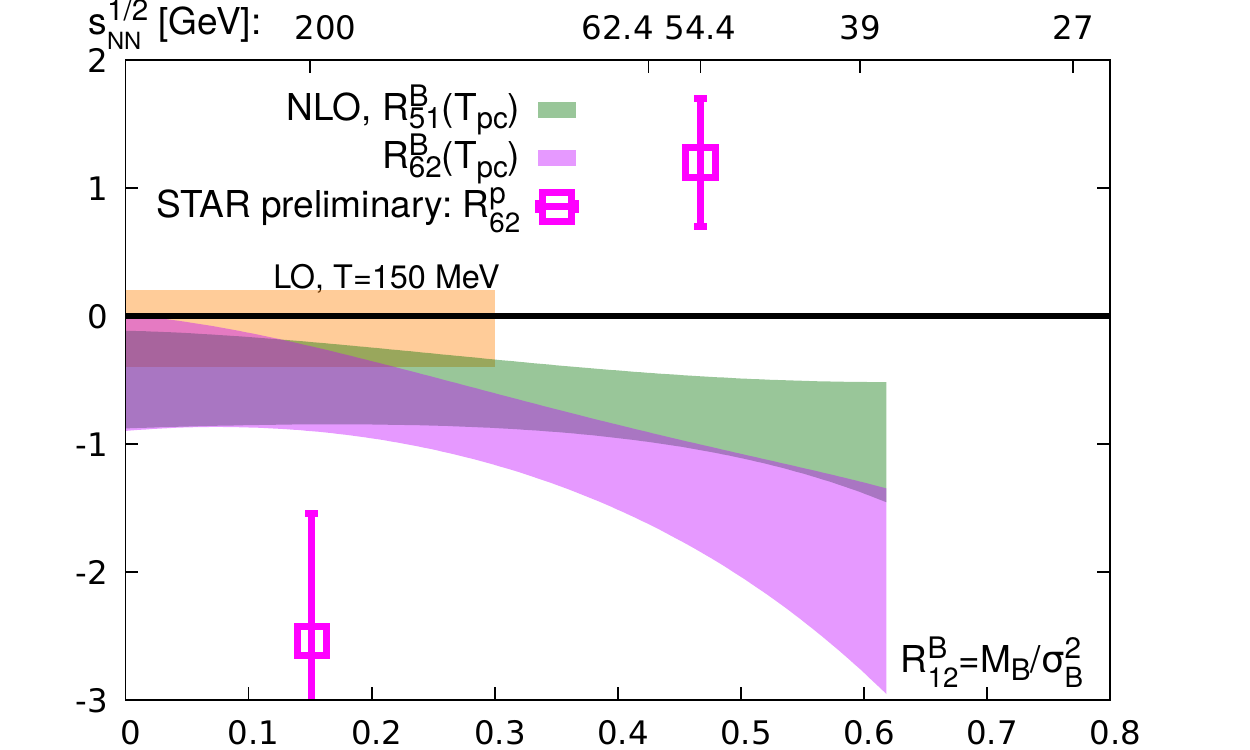}
  \caption{Left figure: Comparison of $R^{B}_{31}(R^{B}_{12})$ and $R^{B}_{42}(R^{B}_{12})$ with corresponding proton fluctuations from STAR. Right figure: Same as left figure but with $R^{B}_{51}(R^{B}_{12})$ and $R^{B}_{62}(R^{B}_{12})$.}
  \label{fig:rnmb}
\end{figure}

In Fig.\ref{fig:rnmb} (left) we show the skewness and kurtosis ratios evaluated along the pseudo-critical transition line $T_{\mathrm{pc}}(\mu_B)$ together with preliminary results of their corresponding net proton-number cumulant ratios measured by STAR \cite{Collaboration2020netproton}\cite{54gev}. To allow for a parameter free comparison, we replace $\mu_B$ on the x-axis with $R_{12}^{B}$ evaluated at $T_{\mathrm{pc}}(\mu_B)$. Of course, many caveats need to be taken into account when comparing baryon-number fluctuations calculated in thermal equilibrium with proton-number fluctuations measured in heavy ion collisions \cite{BEST}. A direct comparison, as shown here, should rather be considered as a first step towards a more thorough analysis. From the comparison, it seems that a freeze-out slightly below $T_{\mathrm{pc}}$ is thermodynamically consistent with the skewness and kurtosis ratios calculated in lattice QCD.

Lastly, we also compute the ratios $R^{B}_{51}(T,\mu_B)$ and $R^{B}_{62}(T,\mu_B)$ which include the hyper-skewness and hyper-kurtosis, respectively,
\begin{align}
  R^{B}_{51}(T,\mu_B)=\frac{S^{H}_{B}\sigma^5_{B}}{M_{B}}=\frac{\chi^{B}_{5}(T,\mu_B)}{\chi^{B}_{1}(T,\mu_B)}, \;\; R^{B}_{62}(T,\mu_B)=\kappa^{H}_{B}\sigma^{4}_{B}=\frac{\chi^{B}_{6}(T,\mu_B)}{\chi^{B}_{2}(T,\mu_B)}.
\end{align}
Since the statistical uncertainties at such high orders are large, we can only use the $N_{\tau}=8$ dataset here. Similar to the skewness and kurtosis ratios, the leading order contributions to both quantities are nearly identical, while the NLO contributions are roughly related by a factor three. Both LO and NLO contributions are negative throughout, such that hyper-skewness and hyper-kurtosis ratios remain negative across the full $\mu_B/T$ region that is being probed. This is especially relevant when comparing to preliminary results of net proton hyper-kurtosis as measured by STAR \cite{54gev}, depicted in Fig.\ref{fig:rnmb} (right). The observed sign change present in the experimental data cannot be explained by thermal QCD at NLO in $\mu_B$. Large NNLO contributions might alter the result at $R^{B}_{12}\simeq0.5$ in favor of the experimental measurement, but results around $R^{B}_{12}\simeq0.15$ would not change much so that it seems impossible to match both data points with equilibrium QCD. We also note that all bands shown in Fig.3 will shift to larger values, when drawn for a temperature below the pseudo-critical temperature, which may be more appropriate, if freeze-out actually takes place below $T_{pc}(\mu_B)$. For orientation we show the LO results for $R_{31}^B\simeq R_{42}^B$ and $R_{51}^B\simeq R_{62}^B$ at $T=150$~MeV and $\mu_B=0$ in Fig.3 (left) and (right), respectively.

\section{Conclusion}

We have calculated up to sixth order cumulant ratios of net baryon number distributions at small $\mu_B$ for strangeness neutral systems, $n_s=0$, with isospin imbalance $n_Q/n_B=0.4$ via (2+1)-flavor HISQ/tree lattice QCD simulations. We found a good agreement when comparing our results for skewness and kurtosis ratios of net baryon-number fluctuations to measurements of net proton-number fluctuations by STAR. The measured skewness and kurtosis ratios are thermodynamically consistent with a freeze-out close-to but smaller than $T_{\mathrm{pc}}(\mu_B)$. We also compared first estimates of fifth and sixth order cumulant ratios to preliminary results obtained by STAR. So far, it seems that the sign change present in the experimental data can not be reproduced by our lattice QCD calculations.

\section{Acknowledgements}
This work was supported by the Deutsche Forschungsgemeinschaft (DFG, German Research Foundation) - project number 315477589 - TRR 211; the German Bundesministerium f\"ur Bildung und Forschung through Grant No. 05P2018 (ErUM-FSP T01) and the European Union H2020-MSCA-ITN-2018-813942 (EuroPLEx). It furthermore received support from the U.S. Department of Energy, Office of Science, Office of Nuclear Physics through (i) the Contract No. DE-SC0012704 and (ii) within the framework of the Beam Energy Scan Theory (BEST) Topical Collaboration, and (iii) the Office of Nuclear Physics and Office of Advanced Scientific Computing Research within the framework of Scientific Discovery through Advance Computing (SciDAC) award Computing the Properties of Matter with Leadership Computing Resources.



%% References


\bibliographystyle{elsarticle-num}
\bibliography{QM19_final.bib}






% \end{thebibliography}

\end{document}